\documentclass[journal]{IEEEtran}
\usepackage{ifpdf}
\usepackage{graphicx}
\usepackage{amssymb}
\usepackage{amsmath}
\usepackage{cite}
\usepackage{epstopdf}
\usepackage{dsfont}
\usepackage[justification=centering]{caption}
\usepackage{color}
\usepackage[linesnumbered,ruled,commentsnumbered,longend]{algorithm2e}

\usepackage{etoolbox}
\usepackage{subcaption}

\newcommand{\mbf}[1]{\mathbf{#1}}
\usepackage{balance}

\newtheorem{definition}{Definition}

\newtheorem{remark}{Remark}

\makeatletter
\def\ScaleIfNeeded{%
\ifdim\Gin@nat@width>\linewidth \linewidth \else \Gin@nat@width \fi
} \makeatother

\begin{document}

\title{Power Minimization for Multi-cell Uplink NOMA with Imperfect SIC}

\author{\IEEEauthorblockN{Ming Zeng, Wanming Hao, Octavia A. Dobre, \emph{Fellow, IEEE}, Zhiguo Ding, \emph{Fellow, IEEE} \\
and H. Vincent Poor,  \emph{Fellow, IEEE}
}
\thanks{M. Zeng is with Laval University, Canada (e-mail: mzeng@mun.ca).

O. A. Dobre is with Memorial University, Canada (e-mail: odobre.mun.ca).}

\thanks{W. Hao is with Zhengzhou University, China (e-mail: wmhao@hotmail.com).}
\thanks{Z. Ding is with The University of Manchester, UK
(e-mail: zhiguo.ding@manchester.ac.uk).}
\thanks{
H. V. Poor is with Princeton University, USA (e-mail: poor@princeton.edu).}
}


\maketitle

\begin{abstract}
In this paper, we investigate a multi-cell uplink non-orthogonal multiple access (NOMA) system with imperfect successive interference cancellation (SIC). {\color{black}The objective of the formulated optimization problem is to minimize the total power consumption under users' quality-of-service constraints}. The considered problem is first transformed into a linear programming problem, upon which centralized and distributed optimal solutions are proposed. Numerical results are presented to verify the {\color{black}performance} of the proposed solutions and
evaluate the {\color{black}impact} of imperfect SIC on the system performance.   
\end{abstract}

\IEEEpeerreviewmaketitle


\section{Introduction}
Recently, non-orthogonal multiple access (NOMA) has received great attention as a promising radio access technology for 5G and beyond networks \cite{RiazBook19, Li_Systems20, Li_WCL20}. In NOMA, users are multiplexed in {\color{black}the} power domain, and thus, the role of power allocation (PA) {\color{black}becomes more important} when compared with orthogonal multiple access (OMA) \cite{RA}. One of the {\color{black}crucial topics on the} PA studies is about spectral efficiency (SE), and existing works have shown that NOMA can achieve higher SE than OMA via appropriate PA{\color{black}\cite{17, 18, Wang_Access19, Khan_IET19}}.

In addition to SE, power consumption is a key criterion. Indeed, power minimization has been considered for various NOMA systems \cite{LLei_CL16, XLi_CL16, JCui_ICC17, LBai_TWC19, YFu_TWC17, ZYang_TWC18, DNi_CL18, 12, JFarah_TVT18}. 
More specifically, the total power minimization problem under quality-of-service (QoS) requirements was studied for single-antenna multi-carrier NOMA systems in \cite{LLei_CL16, XLi_CL16}, which involves a joint allocation of sub-carrier and power. The authors in \cite{LLei_CL16} proved its NP-hardness, and proposed an efficient algorithm via convex relaxation. The authors in \cite{XLi_CL16} first provided the optimal PA under fixed sub-carrier assignments, and on this basis, proposed a low-complexity joint sub-carrier and PA algorithm. 
Additionally, a multiple-input multiple-output (MIMO) NOMA system was studied in \cite{JCui_ICC17}, where the authors proposed a joint PA and receive beamforming scheme. The authors in \cite{LBai_TWC19} further applied vector-perturbation to MIMO-NOMA systems, and proposed a suboptimal joint beamforming and PA strategy. Note that the {\color{black}aforementioned} works focused on single-cell systems. Multi-cell NOMA systems were considered in \cite{YFu_TWC17} and \cite{ZYang_TWC18}, where closed-form power minimization solutions were derived. The authors in \cite{DNi_CL18} further studied the power minimization for a multi-cell multi-carrier NOMA system. A centralized power control algorithm with fixed user assignment was developed first, upon which a greedy user clustering and PA scheme was proposed. In addition, power minimization has also been considered for cognitive radio \cite{12} and {\color{black}distributed-antenna} \cite{JFarah_TVT18} based NOMA systems. 

{\color{black}Note that the above works \cite{LLei_CL16, XLi_CL16, JCui_ICC17, LBai_TWC19, YFu_TWC17, ZYang_TWC18, DNi_CL18, 12, JFarah_TVT18} on power minimization focused on downlink systems. Power allocation is of more significance for uplink systems, since user terminals are power constrained 
\cite{HWang_CL18, Wu_Sensors18, Tweed_Access17, Tweed_ICC18}. 
In this regard, \cite{HWang_CL18} considered a power minimization precoding problem for a MIMO-NOMA uplink system, and proposed a suboptimal joint design of precoders and equalizers. 
Further, \cite{Wu_Sensors18} aimed to minimize the total radio resource consumption cost,
including the cost for the channel usage and that for all sensors' energy consumption under given data delivery requirements. Additionally, \cite{Tweed_Access17} studied outage constrained resource allocation for uplink NOMA, and proposed a suboptimal iterative solution based on successive convex approximation. The result was further extended to a MIMO-NOMA scenario in \cite{Tweed_ICC18}. {\color{black}Note that the works \cite{HWang_CL18, Wu_Sensors18, Tweed_Access17, Tweed_ICC18} mainly focus on single-cell scenarios.}  
}

Research on power minimization for uplink NOMA systems is still at an incipient stage, and more effort is required in this direction. To this end, we consider the power minimization for a multi-cell uplink NOMA system in this paper. In particular, imperfect successive interference cancellation (SIC) is taken into account. {\color{black}This further differentiates our work from \cite{Wu_Sensors18}, which assumes perfect SIC.}
The formulated sum power minimization problem under {\color{black}the users'} QoS requirements is first transformed into a linear programming problem. Then, a closed-form solution is derived, which requires global information at a central entity. To reduce the signalling overhead, a distributed algorithm is further proposed, which is guaranteed to converge to an optimal solution if exists.
{\color{black}This again differentiates our work from \cite{Wu_Sensors18}, which only provides a centralized sub-optimal solution to its formulated problem.}
{\color{black}The presented} numerical results show that: 1) NOMA with perfect SIC outperforms OMA for multi-cell uplink systems; 2) the performance of NOMA deteriorates with the imperfect SIC coefficient, and it can even become worse than OMA if the coefficient is too large; and 3) the proposed distributed solution converges to the optimal solution within a few iterations, yielding substantial signalling overhead savings when compared with the centralized one.

The rest of the paper is organized as follows: Section II introduces the system model and problem formulation; Sections III and IV present the proposed centralized and distributed solutions, respectively; Section IV shows simulation results, whereas Section VI concludes the paper.


\section{System Model and Problem Formulation}
\subsection{System Model}
In this paper, we consider an uplink NOMA system with $M$ cells. Denote the cell set by $\mathcal{M}=\{1, \cdots, M\}$. In cell $m \in \mathcal{M}$, there exists a base station (BS), which serves $N_m$ users using NOMA. Denote the user set in cell $m$ by $\mathcal{N}_m=\{1, \cdots, N_m\}$. The total number of users in the system is then given by $N=\sum_{m=1}^M N_m$.


For $m, m'\in \mathcal{M}$ and $n \in  \mathcal{N}_m$, denote the channel gain between BS $m'$ and user $n$ from BS $m$ by  $h_{n,m,m'}$. 
Let $p_{n,m}$ be the transmit power for user $n$ at cell $m$, satisfying $p_{n,m} \leq p_{n,m}^{\max}$, where $p_{n,m}^{\max}$ is the maximum transmit power. We consider universal frequency reuse among the cells, and thus, {\color{black}each BS receives} inter-cell interference from users in other cells.
The inter-cell interference received by {\color{black}BS} $m$ is given by 
\begin{equation} \label{inter-cell}
I_m= \sum_{m'=1, m' \neq m}^M \sum_{n' \in \mathcal{N}_{m'}} p_{n',m'}h_{n',m',m}.
\end{equation}

Within each cell, users are served using NOMA, and SIC is adopted at the BS to mitigate the inner-cell interference. 
Here we consider that users with better channel gains are decoded first at the BS. Note that the analysis in this paper can be directly applied to other decoding orders. 
Without loss of generality, we assume that users in each cell $m$ are arranged in an ascending order, i.e.,
\begin{equation}
h_{1,m,m} \leq \cdots \leq h_{N_m,m,m}.
\end{equation}

In practice, due to hardware limitation, channel estimation error, low signal quality, etc., decoding error of the interfering signals may
occur. As a result, residual interference may exist after SIC, referred to as imperfect SIC. According to \cite{Wang_Access19, H_T17, H_Sun16, X_Chen18}, the residual interference from imperfect SIC can be modeled as a linear function of the power of the interfering signals, while the coefficient can be attained via long term measurements. 
Then, the inner-cell interference for user $n, n \in \mathcal{N}_m$ is given by{\color{black}\cite{Wang_Access19, H_T17, H_Sun16, X_Chen18}}
\begin{equation} \label{inner-cell}
\hat{I}_{n,m}= \sum_{n' =1}^{n-1} p_{n',m}h_{n',m,m}+ \beta  \sum_{n' =n+1}^{N_m} p_{n',m}h_{n',m,m},  
\end{equation}
where $\beta \in [0,1]$ denotes the coefficient of imperfect SIC, and a higher $\beta$ means more interference.\footnote{{\color{black}Here all the users with residual interference are assumed to use the same coefficient $\beta$ for simplicity. 
Note that different coefficients can also be used for different users. Indeed, the analysis carried in the paper can be generalized to the case with an arbitrary choice of $\beta$ values for the users.}}
{\color{black}On the other hand, when $\beta=1$, the inter-user interference is not cancelled at all. This is clearly the worst case scenario, and the analysis under such a setup can serve as a lower bound on any practical NOMA-based scenario.  
}

Combining \eqref{inter-cell} and \eqref{inner-cell}, the signal-to-interference-plus-noise ratio (SINR) of user $n, n \in \mathcal{N}_m$ is given by
\begin{equation}
\gamma_{n,m}= \frac{p_{n,m}h_{n,m,m}}{\hat{I}_{n,m}+ {I}_{m} +\sigma_m^2 }, 
\end{equation}
where $\sigma_m^2$ denotes the noise power for cell $m$.

\subsection{Problem Formulation}
We aim to minimize the overall transmit power of all users under {\color{black}their target} minimum SINR constraints and maximum transmit power constraints. The considered problem is formulated as {\color{black}follows:}
\begin{subequations} \label{P1}
\begin{align} 
\underset{p_{n,m}}{{\color{black}\min}} & ~  \sum_{m \in \mathcal{M}}\sum_{n \in \mathcal{N}_m} p_{n,m} \\
\text{s.t.}
& ~\gamma_{n,m} \geq \gamma_{n,m}^{\min}, \forall m \in \mathcal{M}, n \in \mathcal{N}_m  \\
& ~0 \leq p_{n,m} \leq p_{n,m}^{\max}, \forall m \in \mathcal{M}, n \in \mathcal{N}_m,
\end{align}
\end{subequations}
where $(\ref{P1}\rm{b})$ and $(\ref{P1}\rm{c})$ denote the minimum SINR constraints and maximum power constraints, respectively.

\section{Centralized Power Control}
In \eqref{P1}, the only non-convex constraint is (\ref{P1}b). To handle it, we rewrite it as follows:
\begin{equation} \label{sinr}
p_{n,m}h_{n,m,m} \geq ( \hat{I}_{n,m}+ {I}_{m} +\sigma_m^2 ) \gamma_{n,m}^{\min}, \forall m \in \mathcal{M}, n \in \mathcal{N}_m
\end{equation}
which is an affine constraint. 

By substituting (\ref{P1}b) with \eqref{sinr}, \eqref{P1} can be re-written as follows: {\color{black}
\begin{subequations} \label{P2}
\begin{align} 
\underset{p_{n,m}}{{\color{black}\min}} & ~  \sum_{m \in \mathcal{M}}\sum_{n \in \mathcal{N}_m} p_{n,m} \\
\text{s.t.}&~\eqref{sinr}, (\ref{P1}\rm{c}).
\end{align}
\end{subequations} }

It can be seen that \eqref{P2} is a linear programming problem, and thus, can be efficiently solved using the simplex method or convex optimization methods, e.g., {\color{black}the} interior-point method. Nonetheless, these numerical methods do not shed light into how the users interact with each other. In the following, we focus on finding a closed-form solution. 


First, we rewrite the transmit power for all users in {\color{black}a} vector form, and define
\begin{equation}
\mbf{p}=[\mbf{p}_1, \cdots, \mbf{p}_m, \cdots, \mbf{p}_M]^T,
\end{equation}
where $\mbf{p}_m=[p_{1,m}, \cdots, p_{n,m}, \cdots, p_{N_m,m}]$ {\color{black}denote} the power values for users in cell $m$. 
Then, the SINR constraints \eqref{sinr} can be re-expressed in matrix form as follows:
\begin{equation} \label{sinr_matrix}
\left[\mathbb{I}_{N \times N}-\mbf{B} \right] \mbf{p}= \mbf{u},
\end{equation}
where $\mathbb{I}_{N \times N}$ is the $N \times N$ identity matrix. Besides, $\mbf{B}$ is an $N \times N$ non-negative matrix, whose $(i, j)$-th element is given by \eqref{Bij} at the top of next page. 

\begin{figure*}[!t]
\begin{align} \label{Bij}
\mbf{B} ({i,j}) &= \begin{cases}
0, & \text{if}~ i=j\\
\frac{\gamma_{n,m}h_{k,m,m}}{h_{n,m,m}},&\text{if}~ i=\sum_{m'=1}^{m-1}N_{m'}+n, j=\sum_{m'=1}^{m-1} N_{m'}+k, k < n \leq N_m \\ 
\frac{\beta \gamma_{n,m}h_{k,m,m}}{h_{n,m,m}},&\text{if}~ i=\sum_{m'=1}^{m-1} N_{m'} +n, j=\sum_{m'=1}^{m-1}N_{m'}+k, n<k \leq N_m  \\
\frac{ \gamma_{n,m}h_{k,m_j,m}}{h_{n,m,m}},&\text{if}~ i=\sum\limits_{m'=1}^{m-1}N_{m'}+n, j=\sum\limits_{m'=1}^{m_j-1}N_{m'}+k, m_j \neq m, n \leq N_m, k \leq N_{m_j} 
\end{cases}
\end{align}
\end{figure*}

Likewise, the $i$-th component of $\mbf{u} $ is given by
\begin{equation}
\mbf{u}(i)=\frac{\gamma_{n,m} \sigma_m^2 }{h_{n,m,m}}, \forall i=\sum_{m'=1}^{m-1}N_{m'}+n, n \leq N_m. 
\end{equation}

$\mbf{B} $ and $\mbf{u}$ are referred to as the normalized interference matrix and noise vector, respectively. It is assumed that $\mbf{B} $ is an irreducible matrix. Based on the Perron-Frobenius Theorem \cite{ESeneta81}, $\mbf{B}$ has a positive real eigenvalue, $\lambda_{\mbf{B}} $, satisfying $\lambda_{\mbf{B}} \geq |\lambda|$ for any eigenvalue $\lambda \neq \lambda_{\mbf{B}} $. $\lambda_{\mbf{B}} $ is called the Perron-Frobenious eigenvalue of $\mbf{B} $, and is associated with strictly positive eigenvectors. According to the Perron-Frobenius Theorem \cite{ESeneta81}, a non-negative solution $\mbf{p}$ to the inequality \eqref{sinr_matrix} exists for any $\mbf{u}\geq 0$, $\mbf{u}\neq 0$, if and only if $\lambda_{\mbf{B}}<1 $. Moreover, the optimal solution $\mbf{p}^\star$ is unique, and can be obtained by 
\begin{equation} \label{PA solution}
\mbf{p}^\star= \left[\mathbb{I}_{N \times N}-\mbf{B} \right]^{-1} \mbf{u}.
\end{equation}

Note that to determine the feasibility of problem \eqref{P2}, we need to consider the maximum power constraint $(\ref{P1}\rm{c})$ as well. Under the condition that a non-negative $\mbf{p}^\star$ exists, \eqref{P2} is feasible if and only if $\mbf{p}^\star (i) \leq p_{n,m}^{\max}, i=\sum_{m'=1}^{m-1}N_{m'}+n, n \leq N_m $ holds for all users.   

\begin{remark}
It has been shown that the sum rate maximization problem is NP-hard when inner-cell interference exists \cite{Luo_JSAC08}. In contrast, the corresponding sum power minimization problem {\color{black}can be} solvable as shown above. This difference can be explained by the monotonicity. For the sum rate maximization problem, increasing the rate of one user often yields more interference to other users, thus lowering their rates. As a result, the monotonicity is unclear, and how to balance the rates among users is complicated. For the {\color{black}considered} power minimization problem, lowering the power of one user also reduces the interference to other users, thus lowering their power requirement. As a result, the monotonicity exists, which limits each user to transmit with the minimum required power.  
\end{remark}

\section{Distributed Power Control}
The centralized power control requires global information at the BSs or a central entity, which involves a large amount of signalling overhead, especially for large-scale multi-cell networks. To reduce the signalling overhead, we propose a distributed iterative power control algorithm, which only requires local information at each BS. 

Specifically, the proposed distributed iterative power control algorithm works as follows: we first set the power values for all users to zero, i.e., $\mbf{p}=\mbf{0}$; then, we update the power values for the cells one by one based on \eqref{PA solution_m} given below, while keeping the power values for other cells fixed; this procedure is repeated until the SINR constraints for all users are satisfied under given threshold or the power constraint for any user is violated. Note that whenever the latter case occurs, the considered problem is infeasible.  
The pseudo-code is summarized in Algorithm~\ref{DIPCA}. 

Now we show in detail how each cell updates the power values for its associated users. Without loss of generality, we consider cell $m$. The power update is conducted when the power values for users in other cells are fixed. As a result, $I_m$ is fixed and known. 

Similar to \eqref{sinr_matrix}, the SINR constraints \eqref{sinr} for users in cell $m$ can be rewritten in {\color{black}a} matrix form as follows:
\begin{equation} \label{sinr_matrix_m}
\left[\mathbb{I}_{N_m \times N_m}-\mbf{B}_m \right] \mbf{p}_m^T= \mbf{u}_m,
\end{equation}
where $\mathbb{I}_{N_m \times N_m}$ is the $N_m \times N_m$ identity matrix. Besides, $\mbf{B}_m$ is an $N_m \times N_m$ non-negative matrix, whose $(i, j)$-th component is given by
\begin{align}
\mbf{B}_m ({i,j}) &= \begin{cases}
0, & \text{if}~ i=j\\
\frac{\gamma_{i,m}h_{j,m,m}}{h_{i,m,m}},&\text{if}~ i> j \\
\frac{\beta \gamma_{i,m}h_{j,m,m}}{h_{i,m,m}},&\text{if}~ i< j
\end{cases}.
\end{align}

Likewise, the $i$-th component of $\mbf{u} $ is given by
\begin{equation}
\mbf{u}_m(i)=\frac{\gamma_{i,m} ({I}_{m}+ \sigma_m^2)}{h_{i,m,m}}. 
\end{equation}

Accordingly, the unique optimal solution $\mbf{p}_m^\star$ is given by
\begin{equation} \label{PA solution_m}
{\mbf{p}_m^\star}^T= \left[\mathbb{I}_{N_m \times N_m}-\mbf{B}_m \right]^{-1} \mbf{u}_m.
\end{equation}

\subsection{Convergence}
Here we only consider the case when the considered problem is feasible. 
Before showing the convergence of the proposed algorithm, we first introduce the following definition:
\begin{definition}
The interference function $\mbf{I(p)}$ is standard if for all $\mbf{p} \geq 0$, the following properties hold:
\begin{itemize}
\item Positivity: $\mbf{I(p)}>0$;
\item Monotonicity: if $\mbf{p} \geq \mbf{p'}$, then $\mbf{I(p)}>\mbf{I(p')}$;
\item Scalability: for all $\alpha>1$, $\alpha \mbf{I(p)}>\mbf{I(\alpha p)}$.
\end{itemize}
\end{definition}

Note that $\mbf{I(p)}=\left(\mbf{I}_1\mbf{(p)}, \cdots,  \mbf{I}_N\mbf{(p)} \right)$ is a vector-valued function, with $\mbf{I}_i\mbf{(p)}$ denoting the effective interference that user $i$ must overcome.
Since both the inter-cell and inner-cell interferences considered in this paper are affine, it can be easily verified that the corresponding interference function satisfies the three properties of standard interference function. 
Therefore, the corresponding interference function is standard, and according to \cite[Th. 2]{RDYates_JSAC95}, the proposed algorithm is guaranteed to converge to the optimal solution.  

\subsection{Distributed Implementation}
The proposed algorithm is distributed since each BS $m$ only requires the following two kinds of local information:
\begin{itemize}
\item the channel gains between each BS and its associated users;
\item the interference plus noise power at each BS.
\end{itemize}

It is clear that such information can be {\color{black}acquired} within cell $m$. On this basis, the power control at BS $m$ can be performed locally according to \eqref{PA solution_m}.

\begin{algorithm}[t]
\caption{{\small{Proposed Distributed Iterative Power Control Algorithm (DIPCA).}}} \label{DIPCA}
{\bf{Initialize}} $\varepsilon^{\star} \leftarrow 10^{-6}$; $\varepsilon \leftarrow 1$; ${\bf{p}}^{(0)} \leftarrow \mbf{0}$; $\rm{indicator} \leftarrow 0$; $l \leftarrow 1$; \\ 
\While{$\varepsilon \geq \varepsilon^{\star}$}
{
\For{$m \leftarrow 1: M$}
 {
 $\mbf{p}_m^{(l)}  \leftarrow \left[\mathbb{I}_{N_m \times N_m}-\mbf{B}_m \right]^{-1} \mbf{u}_m$; \\
 \For{$n \leftarrow 1: N_m$}
 { \If{$\mbf{p}_m^{(l)}(n)> p_{n,m}^{\max}$ \rm{or} $\mbf{p}_m^{(l)}(n)< 0$}
    {$\rm{indicator} \leftarrow 1$; \\
    break;
    }
 }
 } 
 \If{$\rm{indicator}=1$}
    {break;
    }
 $\varepsilon \leftarrow |\mbf{p}^{(l)}-\mbf{p}^{(l-1)}|_2^2$; \\
 $l \leftarrow l+1$;
}
\end{algorithm}


\section{Simulation Results}
In this section, simulations are conducted to verify the accuracy of the proposed solutions, and evaluate the effect of imperfect SIC on the system performance. The default simulation parameters are as follows: 
There are three cells, each with three users, namely $M=3$ and $N_m=3$. The cell radius is 100 m, and users are randomly uniformly distributed within each cell. The pathloss model is $30.6+36.7\log_{10}(d)$, where $d$ is the distance in m. The system bandwidth is 10 MHz, while the noise power spectral density is $-174$ dBm/Hz. The maximum power constraint for each user is $30$ dBm.  
Unless mentioned explicitly, results are averaged over $10^4$ random trials.

To show how the imperfect SIC coefficient $\beta$ affects the performance of NOMA systems, we adopt four different $\beta$ values, i.e., $\beta=0,~0.05,~0.1$ and $0.15$. 
Moreover, we also adopt OMA with equal degrees of freedom for users within the same cell as a baseline.
Due to orthogonality within each cell, users only receive inter-cell interference. The optimal solution can be obtained based on \eqref{PA solution} as well, after modifying matrices $\mbf{B}$ and $\mbf{u}$.

\begin{figure} 
\centering
\includegraphics[width=0.45\textwidth]{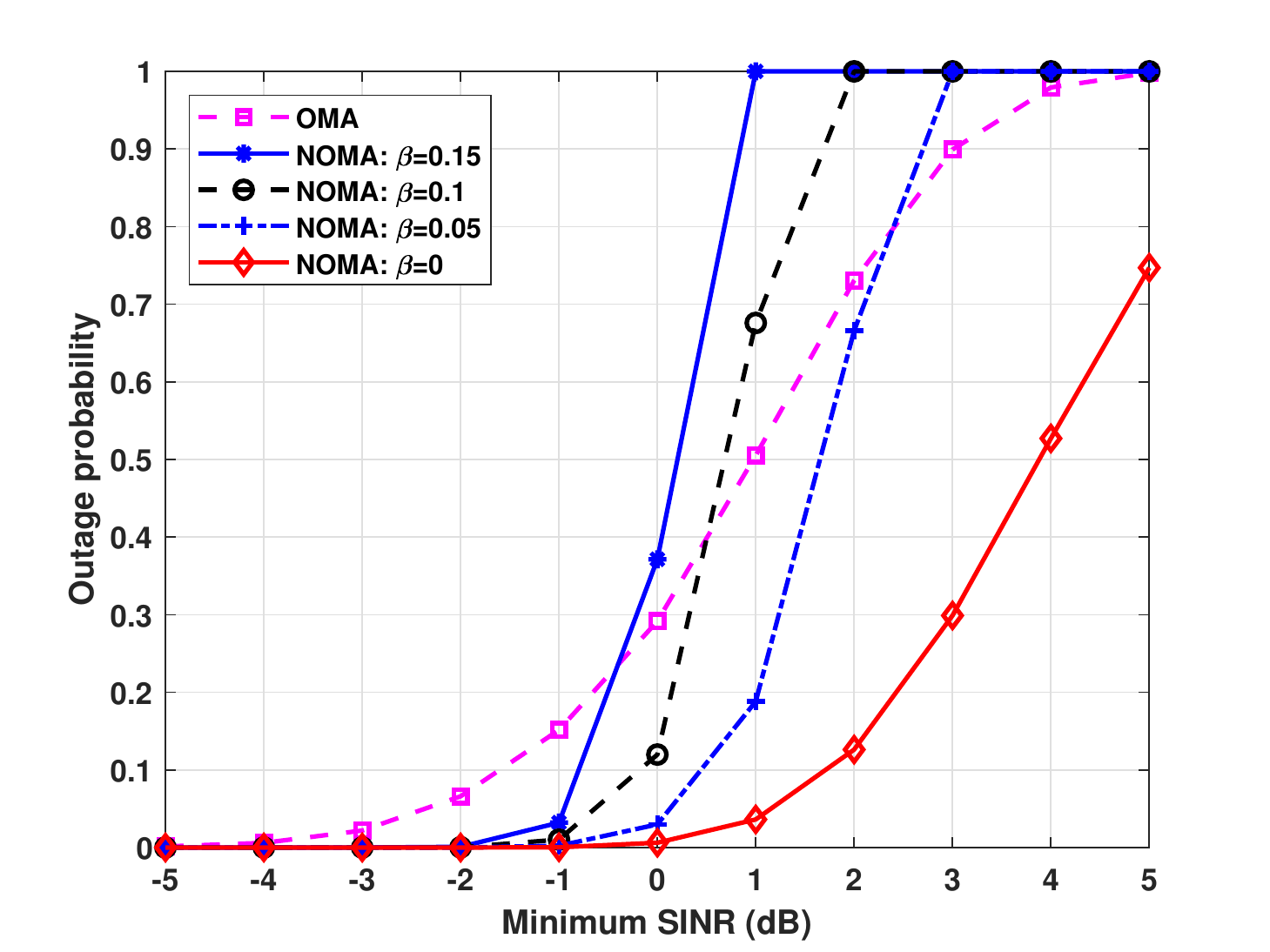}
\caption{Outage probability versus minimum SINR requirement $\gamma_{n,m}^{\rm{min}}$ for each user.} \label{fig_2}
\end{figure}

\begin{figure} 
\centering
\includegraphics[width=0.45\textwidth]{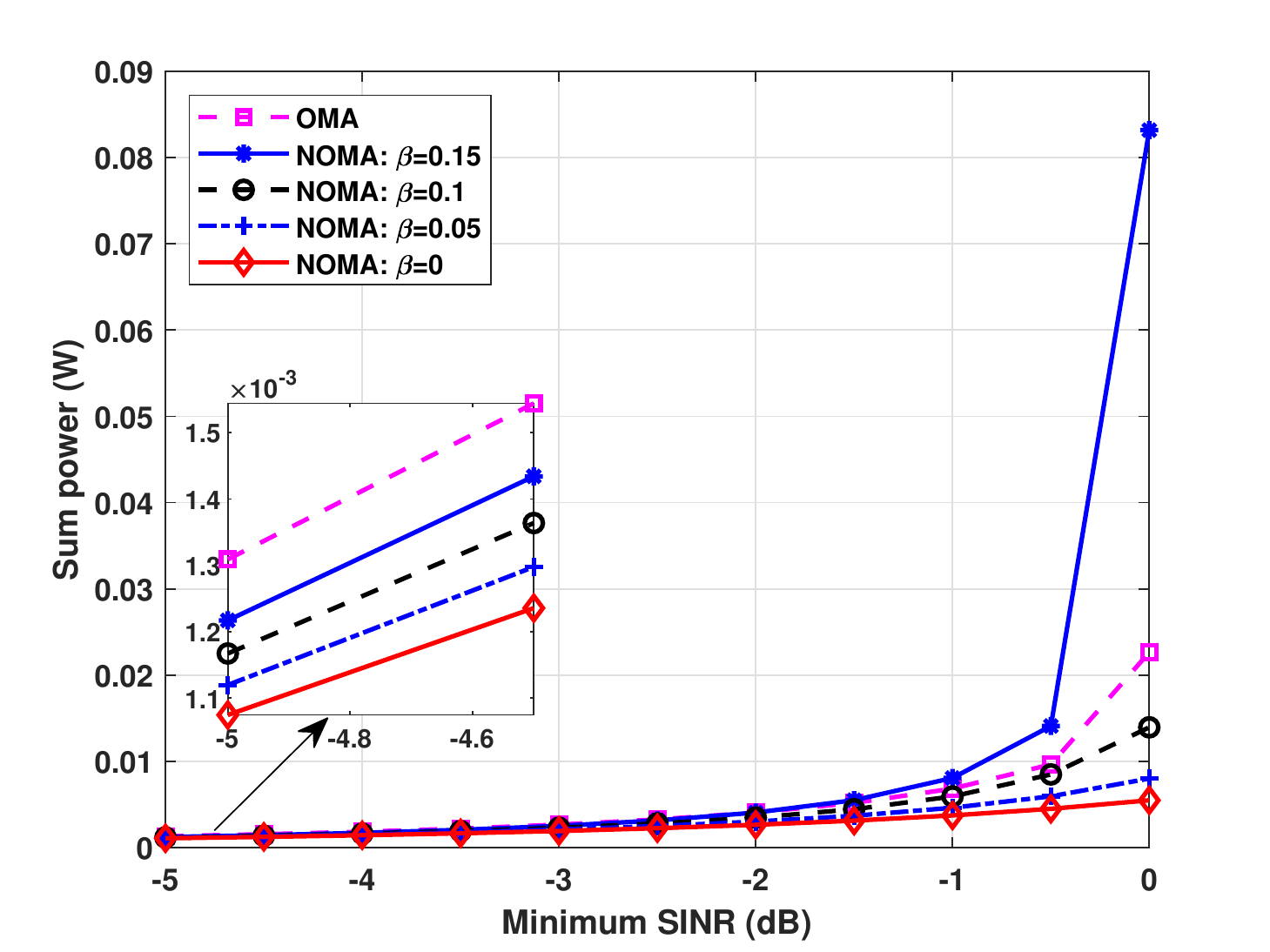}
\caption{Sum power versus minimum SINR requirement $\gamma_{n,m}^{\rm{min}}$ for each user.} \label{fig_1}
\end{figure}

{\color{black}
In simulation, it was found that the obtained results from the centralized solution, distributed solution and interior-point method are the same for NOMA, which shows the accuracy of the proposed centralized and distributed solutions. Accordingly, we only use results from the centralized power control as a representative for NOMA in Figs.~\ref{fig_2} and ~\ref{fig_1}.
More specifically, Fig.~\ref{fig_2} illustrates how the outage probability varies with the users' minimum SINR requirement for both NOMA and OMA. Note that outage occurs whenever the considered problem is infeasible, i.e., there exist no power values satisfying the constraints.
It is clear that NOMA with perfect SIC, i.e., $\beta=0$ achieves lower outage probability than OMA under any minimum SINR requirement, which shows the superiority of NOMA over OMA. Besides, NOMA with $\beta=0.05, 0.1, 0.15$ also outperforms OMA under low minimum SINR requirement, but becomes inferior to OMA under high minimum SINR requirement. This is because under such conditions, the residual inter-user interference from imperfect SIC is severe, which leads to high outage probability.


Figure~\ref{fig_1} shows how the required sum power varies with the users' minimum SINR requirements for both NOMA and OMA. For a fair comparison, only the trials that lead to feasible solutions for all minimum SINR requirements under both NOMA and OMA are included. Meanwhile, only the results under $\gamma_{n,m}^{\rm{min}} \leq 0$ dB are considered, since the feasible trial set under $\gamma_{n,m}^{\rm{min}} \geq 1$ is empty according to Fig.~\ref{fig_2}.
As expected, the required sum power increases with the minimum SINR requirement for both NOMA and OMA. Meanwhile, NOMA with perfect SIC outperforms OMA under any minimum SINR requirement, which again illustrates the superiority of NOMA over OMA. However, the required sum power grows with $\beta$ under NOMA, due to increased residual inter-user interference from imperfect SIC. In particular, NOMA with $\beta=0.15$ consumes more power than OMA once $\gamma_{n,m}^{\rm{min}} \geq -1.5$ dB. 
}

\begin{figure} 
\centering
\includegraphics[width=0.45\textwidth]{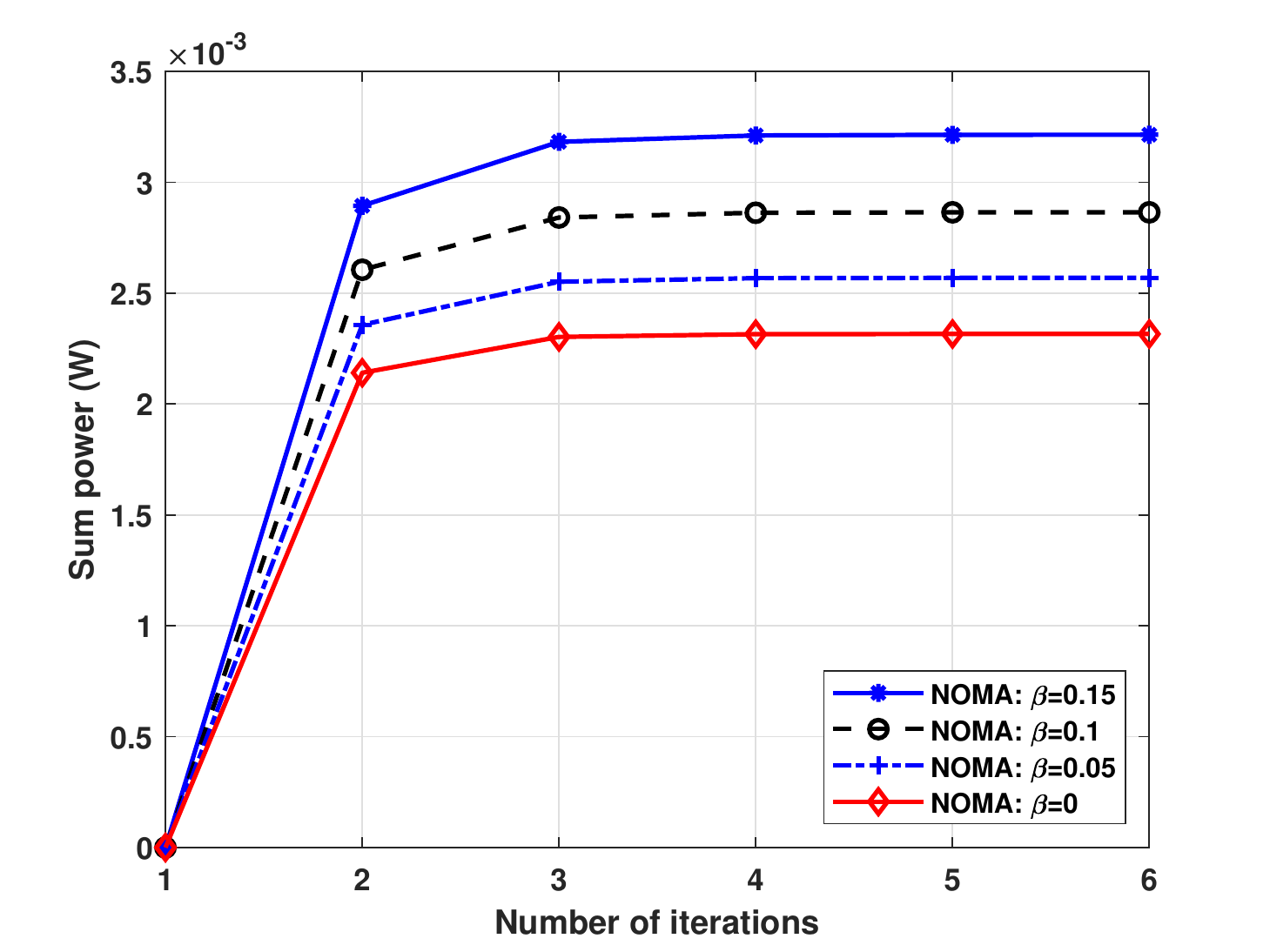}
\caption{Number of iterations required for the distributed power control algorithm to converge.} \label{fig_3}
\end{figure}

Figure~\ref{fig_3} plots the number of iterations required for the proposed distributed iterative power control algorithm to converge. Here only a single trial is considered, and the minimum SINR requirement is set to -2.5 dB. 
It can be seen that after the initial allocation, only five iterations are needed for the proposed distributed algorithm to converge, which shows its effectiveness. Note that an iteration here means that all cells update their power values {\color{black}sequentially}.


\section{Conclusion}
In this paper, the sum power minimization problem has been investigated for a multi-cell uplink NOMA system under {\color{black}the} minimum QoS constraint for each user. Universal frequency reuse and imperfect SIC were assumed, which lead to inter-cell and inner-cell interference, respectively. The formulated problem was first transformed into be a linear programming problem, and further, centralized and distributed solutions were derived. Presented numerical results showed that NOMA with perfect SIC outperforms OMA in terms of sum power consumption and outage probability. However, the gain declines as the imperfect SIC coefficient $\beta$ increases, and even disappears when $\beta$ is large enough. Additionally, the proposed distributed solution converges to the optimal solution within a few iterations, which can yield substantial signalling overhead savings when compared with the centralized one.  

\bibliographystyle{IEEEtran}
\balance
\bibliography{IEEEabrv,conf_short,jour_short,mybibfile}

\end{document}